\documentclass[prc,twocolumn,showpacs,preprintnumbers,amsmath,amssymb,superscriptaddress,floatfix,nofootinbib]{revtex4}
\usepackage{graphicx}%

\newcommand{\Slash}[1]{\ooalign{\hfil/\hfil\crcr$#1$}}

\begin{document}
\title {The role of the $N^*(2080)$ resonance in the $\vec{\gamma} p \to K^+
\Lambda(1520)$ reaction}
\author{Ju-Jun Xie} \email{xiejujun@ific.uv.es}
\affiliation{Instituto de F\'\i sica Corpuscular (IFIC), Centro Mixto
CSIC-Universidad de Valencia, Institutos de Investigaci\'on de
Paterna, Aptd. 22085, E-46071 Valencia, Spain}
\affiliation{Department of Physics, Zhengzhou University, Zhengzhou,
Henan 450001, China}
\author{J. Nieves}
\affiliation{Instituto de F\'\i sica Corpuscular (IFIC), Centro Mixto
CSIC-Universidad de Valencia, Institutos de Investigaci\'on de
Paterna, Aptd. 22085, E-46071 Valencia, Spain}
\date{\today}%

\begin{abstract}
We investigate the $\Lambda(1520)$ photo-production in the
$\vec{\gamma} p \to K^+ \Lambda(1520)$ reaction within the effective
Lagrangian method near threshold. In addition to the "background"
contributions from the contact, $t-$channel $K$ exchange, and
$s-$channel nucleon pole terms, which were already considered in
previous works, the contribution from the nucleon resonance
$N^*(2080)$ (spin-parity $J^P = 3/2^-$) is also considered.  We show
that the inclusion of the nucleon resonance $N^*(2080)$ leads to a
fairly good description of the new LEPS differential cross section
data, and that these measurements can be used to determine some of
the properties of this latter resonance. However, serious
discrepancies appear when the predictions of the model are compared
to the photon-beam asymmetry also measured by the LEPS
Collaboration.

\end{abstract}
\pacs{13.75.Cs.; 14.20.-c.; 13.60.Rj.} \maketitle

\section{Introduction}

The $\Lambda(1520)$ ($\equiv \Lambda^*$) photo-production in the
$\gamma p \to K^+ \Lambda^* $ reaction is an interesting tool to
gain a deeper understanding of the interaction among strange hadrons
and also on the nature of baryon resonances. There have been some
experimental efforts dedicated to this reaction. There exist
measurements from two old experiments in the high energy region
$E_{\gamma} = 2.8 - 4.8$ GeV by the LAMP2
Collaboration~\cite{barber80}, and $E_{\gamma} = 11$ GeV by Boyarski
et al.~\cite{boyarski71}. Recently, this reaction has been examined
at photon energies below $2.4$ GeV in the SPring-8 LEPS
experiment~\cite{leps1,leps2}. For an invariant $\gamma p$ mass $W
\simeq 2.11$ GeV, this latter experiment has reported a new bump
structure in the differential cross section at forward $K^+$ angles,
which might hint to a sizeable contribution from nucleon resonances
in the $s-$channel.

On the theoretical side, there exist several effective hadron
Lagrangian studies of this reaction for laboratory photon energies
ranging from threshold ($\approx $ 1.7~GeV), up to about 5 GeV,
where the old experimental measurements were available. Different
dominant mechanisms have been proposed to describe the LAMP2 high
energy results for angular and energy differential cross sections.
Thus, in Refs.~\cite{titovprc7274,sibiepja31} it is claimed a large
contribution from the $t-$channel $K^*$ exchange, while a large
contribution from a gauge invariant contact term along with
$t-$channel $K$ exchange and baryon pole ($s-$ and $u-$ channels) is
advocated in ~\cite{nam1,nam2}. On the other hand, a quark-gluon
string reaction mechanism, which was realized in the $K$ meson
Reggeon exchange model, is also able to reproduce the available
experimental data in the high photon energy region from $E_{\gamma}
\simeq 2.8$ GeV up to $5$ GeV, as discussed in~\cite{toki}. In this
latter work, the coupling of the $\Lambda^*$ resonance, which is
dynamically generated within a chiral unitary
model~\cite{GarciaRecio:2005hy}, to the $N\bar{K^*}$ pair is shown
to be quite small and hence, the contribution of $t-$channel $K^*$
exchange is taken to be much smaller than that due to exchange of
the pseudoscalar $K$ meson. Indeed, this minor role played by the
$K^*$ exchange gets support from the decay asymmetry measured by the
CLAS Collaboration~\cite{claselectro}. Besides, the photon beam
asymmetry for $K \Lambda^*$ photo-production has been also proposed
in Ref.~\cite{Nam:2006cx} as tool to unravel the reaction mechanism,
since it is predicted to be very small for the contact and $s-$ and
$u-$ baryon pole terms, while much larger for $t-$channel $K$ and
$K^*$ exchanges. Moreover, $K$ and $K^*$ exchanges give rise to
different signs of the photon-beam asymmetry~\cite{Nam:2006cx}.

All these models fail, however, to describe the forward bump
structure that appears in the new LEPS differential cross section
data at low energies.

In the present work, we reanalyze the $\vec{\gamma} p \to K^+
\Lambda^* $ reaction within the effective Lagrangian method
near-threshold, where the new experimental data from the LEPS
Collaboration~\cite{leps2} have been taken. In addition to the
"background" contributions from the contact, $t-$channel $K$
exchange, and $s-$channel nucleon pole terms, which were already
considered in previous works, we have also studied possible
contributions from nucleon resonances aiming at describing the bump
at forward angles reported by LEPS. Unfortunately, the information
about nucleon resonances in the relevant mass region ($\sim 2.1$)
GeV is scarce~\cite{pdg2008}, which means that the evidence of their
existence and the knowledge of their properties are poor. Thus, it
is necessary to rely on theoretical schemes, such that of
Ref.~\cite{simonprd58,Capstick:1992uc} based in a quark model (QM)
for baryons. Among the possible nucleon resonances, we have finally
considered only the two-star $D-$wave $J^P=3/2^-$ $N^*(2080)$
($\equiv N^*$) one, which is predicted to have visible
contributions~\cite{simonprd58} to the $\gamma p \to K^+ \Lambda^* $
reaction. Although the $N^*(2080)$ resonance is listed in the
Particle Data Group (PDG) book, the evidence of its existence is
poor or only fair and further work is required to verify their
existence and to know its properties, accordingly, its total decay
width and branching ratios are not experimentally known, either. In
this respect, we show in this work how the LEPS measurements could
be used to determine some of the properties of this resonance.

To end this introduction, we would like to mention that in
Refs.~\cite{nam2,nam3}, the role played by the $N^*(2080)$ resonance
in the $\vec{\gamma} p \to K^+ \Lambda^* $ reaction has been also
studied.  In these works, in sharp contrast with our findings, it is
pointed out that the $N^*(2080)$ resonance has a negligible
contribution and therefore its contribution is not expected to
explain the bump structure at forward angles reported by LEPS. This
is greatly due to the small $N^*(2080)\Lambda^* K^+$ coupling and
large width of the resonance used in Refs.~\cite{nam2,nam3}, and we
will comment on this below.

The paper is organized as follows. In Sect.~\ref{sec:formalism}, we
shall discuss the formalism and the main ingredients of the model, while our
results and conclusions are presented in Sects.~\ref{sec:results} and
~\ref{sec:conclusions}.
%
%%%%%%%%%%%%%%%%%%%%%%%%%%%%%%%%%%%%%%%%%%%%%%%%%%%%%%%%%%%%%%%%%%
%
\section{Formalism and ingredients} \label{sec:formalism}

\begin{figure}[htdp]
\includegraphics[scale=0.5]{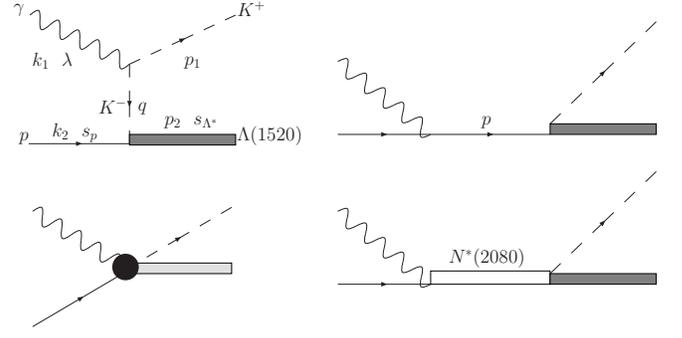}
\vspace{-0.2cm} \caption{Model for the $\gamma p \to \Lambda^* K^+$
reaction. It consists of $t-$channel $K$ exchange, $s-$channel nucleon
and nucleon resonance, and contact terms. In the first diagram, we also
show the definition of the kinematical ($k_1,k_2,p_1,p_2$) and
polarization variables ($\lambda, s_p,s_{\Lambda^*}$) that we used in
the present calculation. In addition, we use
$q=k_1-p_1$}. \label{diagram}
\end{figure}

The basic tree level Feynman diagrams for the
$\vec{\gamma} p \to K^+ \Lambda^* $
reaction are depicted in the Fig.~\ref{diagram}. These include the
$t-$channel $K$ exchange, $s-$channel nucleon  and nucleon
resonance, and contact terms. To compute the contributions of these
terms, we use the  interaction Lagrangian
densities of Refs.~\cite{toki,nam2},
\begin{eqnarray}
\mathcal{L}_{\gamma KK} &=& -ie (K^- \partial^{\mu} K^+ -
K^+\partial^{\mu}  K^- ) A_{\mu}, \\ \label{eq:gamakk}
\mathcal{L}_{Kp\Lambda^*} &=&
\frac{g_{KN\Lambda^*}}{m_{K}}\bar{\Lambda}^{*\mu} (\partial_{\mu}
K^-){\gamma}_{5}p\,+{\rm h.c.},  \label{eq:knl} \\
\mathcal{L}_{\gamma pp} &=&
-e\bar{p}\left(\Slash{A}-\frac{\kappa_{p}}{2M_{N}}
\sigma_{\mu\nu}(\partial^{\nu}A^{\mu})\right) p + {\rm h.c.}, \\
\mathcal{L}_{\gamma Kp\Lambda^*} &=&
-ie\frac{g_{KN\Lambda^*}}{m_{K}}\bar{\Lambda}^{*\mu} A_{\mu}
K^-{\gamma}_{5}p\,+{\rm h.c.}, \\ \label{eq:eqfin}
\mathcal{L}_{\gamma N N^*} &=& \frac{ie f_1}{2m_{N}}
\bar{N}^*_{\mu} \gamma_{\nu} F^{\mu \nu} N \, - \nonumber \\
&&\frac{e f_2}{(2m_{N})^2}\bar{N}^*_{\mu} F^{\mu \nu} \partial_{\nu}
N\,+{\rm h.c.}, \\ \label{eq:eqgamaN}
\mathcal{L}_{K \Lambda^* N^*}
&=& \frac{g_1}{m_{K}} \bar{\Lambda}^*_{\mu} \gamma_{5}
\gamma_{\alpha}
(\partial^{\alpha} K)
N^{* \mu} \, +  \nonumber \\
&& \frac{i g_2}{m_{K}^2} \bar{\Lambda}^*_{\mu} \gamma_5
\left (\partial^{\mu} \partial_{\nu} K\right)  N^{*\nu} \,+{\rm h.c.}, \label{eq:eqknstar}
\end{eqnarray}
where $e=\sqrt{4\pi\alpha} > 0$ ($\alpha = 1/137.036$ is the
fine-structure constant), $\kappa_p=1.79$, $A_\mu$ and $F_{\mu \nu} =
\partial_{\mu}A_{\nu}-\partial_{\nu}A_{\mu}$ are the proton charge and
magnetic moment, and the photon
field and electromagnetic field tensor, respectively. We use the
Rarita-Schwinger formalism~\cite{rarita,nath} to describe the
spin $J=3/2$ $\Lambda^*$ and $N^*$ resonances, while the $N^*(2080)$
electromagnetic $f_{1,2}$ and hadronic $g_{1,2}$ couplings will be
discussed below.

With the effective interaction Lagrangian densities given above,
we can easily construct the invariant scattering amplitudes,
\begin{equation}
-iT_i=\bar u_\mu(p_2,s_{\Lambda^*}) A_i^{\mu \nu} u(k_2,s_p)
\epsilon_\nu(k_1,\lambda)
\end{equation}
where $u_\mu$ and $u$ are dimensionless Rarita-Schwinger and Dirac
spinors, respectively, while $\epsilon_\nu(k_1,\lambda)$ is the
photon polarization vector. The reduced  $A_i^{\mu\nu}$ amplitudes read.
\begin{eqnarray}
A_t^{\mu\nu} &=& -e\frac{g_{KN\Lambda^*}}{m_{K}} \frac{1}{q^2-m_K^2}
q^\mu (q^\nu - p_1^\nu)\gamma_5\, f_{\rm c}, \label{eq:at} \\
A_s^{\mu\nu} &=& -e\frac{g_{KN\Lambda^*}}{m_{K}} \frac{1}{s-M_N^2}\,
p_1^\mu \gamma_5 \big \{\Slash k_1 \gamma^\nu \, f_{\rm s} +(\Slash
k_2+M_N) \gamma^\nu \, f_{\rm c}  \nonumber \\
&& + (\Slash k_1 +\Slash k_2+M_N) {\rm i} \, \frac{\kappa_p}{2M_N}\sigma_{\nu\rho}k_1^\rho\, f_{\rm s}\big\}, \label{eq:as} \\
A_c^{\mu\nu} &=& e\frac{g_{KN\Lambda^*}}{m_{K}}  g^{\mu\nu}
\gamma_5\, f_{\rm c}, \label{eq:ac} \\
A_{R}^{\mu\nu} &=& \gamma_5 (\frac{g_1}{m_K}\Slash p_1 g^{\mu \rho}
- \frac{g_2}{m^2_K}  p_1^{\mu} p_1^{\rho}) \frac{\Slash k_1 + \Slash
k_2 + M_{N^*}}{s - M_{N^*}^2+iM_{N^*}\Gamma_{N^*}} \nonumber
\\ && P_{\rho \sigma}
\Big (\frac{ef_1}{2m_N}(k_1^{\sigma}\gamma^{\nu}-g^{\sigma\nu}\Slash k_1)
+ \frac{ef_2}{(2m_N)^2} \nonumber \\
&& (k_1^{\sigma}k_2^{\nu}-g^{\sigma\nu}k_1 \cdot k_2)\Big)f_R, \label{eq:ar}
\end{eqnarray}
The sub-indices $t, s, c$ and $R$ stand for the $t-$channel kaon
exchange, $s-$channel nucleon pole, contact, and resonance $N^*$ pole
terms\footnote{We do not consider in this work the $t-$channel $K^*$
exchange and the $u$-channel hyperon pole terms, since we expect these
contributions to be small, as it was discussed in Ref.~\cite{toki}.},
and
\begin{eqnarray}
P_{\rho \sigma} &=& -g_{\rho \sigma} +
\frac{1}{3}\gamma_{\rho}\gamma_{\sigma}
+\frac{2}{3M^2_{N^*}}(k_1+k_2)_{\rho}(k_1+k_2)_{\sigma} \nonumber \\
&&
+\frac{1}{3M_{N^*}}(\gamma_{\rho}(k_1+k_2)_{\sigma}-\gamma_{\sigma}(k_1+k_2)_{\rho}).
\end{eqnarray}
Note, that there exist some ambiguities when dealing with the
propagation and couplings of spin 3/2 off-shell
particles~\cite{Mizutani:1997sd, Pascalutsa:1999zz, Pascalutsa:2000kd}. In this
exploratory work, we ignore this problem, since we are using a tree
level approach and possible effects might be partially effectively
encoded into the phenomenological $N^*$ hadron couplings, which will
be fitted to data. This interesting issue deserves future
detailed research.  Besides, $M_{N^*}$ and $\Gamma_{N^*}$ are the
mass and the total decay width of the $N^*$ resonance. We have
examined here two different scenarious. In the first one, we set
$M_{N^*}=2.08$ GeV, as quoted in the PDG~\cite{pdg2008}. On the other
hand, since $\Gamma_{N^*}$ has a large experimental
uncertainty~\cite{pdg2008}, in this first scenarious, we use
$\Gamma_{N^*} = 300$ MeV for the numerical calculations.  These values
have been also used in Ref.~\cite{yongprc77}, where the contributions
of the $N^*$ resonance in the $\gamma p \to K\Sigma(1385)$ reaction
were examined, and found to give important contributions to this
reaction.  In the second scenarious, we have fitted both, the mass and
the total width of the $N^*$ resonance to the LEPS differential cross
section data.

Up to this point, the $T$-matrix is gauge invariant. However, we ought
to introduce the compositeness of the hadrons. This is usually
achieved by including form-factors in the amplitudes in such manner
that gauge invariance is preserved\footnote{For the sake of brevity
and to avoid repeating similar equations in
Eqs.~(\ref{eq:at})-(\ref{eq:ar}) we have already included form-factors
($f_s,f_c,f_R$). Details are given in what follows.}.  There is no an
unique theoretical way to introduce the form-factors, this was
discussed at length in the late nineties~\cite{ohta89,
haberzettl98,Davidson:2001rk, Janssen:2001wk}. We adopt here the
scheme used in the previous works \cite{titovprc7274,nam1,nam2}, where
the prescription of Ref.~\cite{haberzettl98} was used. We take the
following parameterization for the four-dimensional form-factors
\begin{eqnarray}
f_{\rm i} &=&\frac{\Lambda^4_i}{\Lambda^4_i+(q_{\rm i}^2-M_{\rm i}^2)^2},
\quad {\rm i}={\rm s,t, R} \\
f_{\rm c} &=& f_{\rm s}+f_{\rm t}-f_{\rm s}f_{\rm t}, \quad {\rm
and} \left\{\begin{array}{l}                               q_{\rm
  s}^2=q_{\rm R}^2=s,\, q_{\rm t}^2= q^2, \cr
M_{\rm s} = M_N, \cr
 M_{\rm R} = M_{N^*}, \cr
M_{\rm t} = m_K.
\end{array}\right. \label{F1}
\end{eqnarray}
We will consider different cut-off values for the background and
resonant terms, i.e.  $\Lambda_s=\Lambda_t \ne \Lambda_R$.

In the expressions of the different contributions to the $T$
amplitude, given in Eqs.~(\ref{eq:at})-(\ref{eq:ar}), we have
already included the form-factors.  The form of $f_{\rm c}$ is
chosen such that the on-shell values of the coupling constants are
reproduced. For the sake of simplicity, we neglect the terms
affected by the $f_s$ form-factor\footnote{Those terms are
gauge-invariant by themselves.}, since they are greatly suppressed
by it~\cite{nam1,nam2}.

We take $g_{KN\Lambda^*}=10.5$, as determined from the $\Lambda^*
\to p K^-$ decay width (we use for the full decay width
$\Gamma_{\Lambda^*} = 15.6$ MeV and a value of 0.45 for the
$\Lambda^* \to \bar{K}N$ branching ratio~\cite{pdg2008}), while the
$N^* N \gamma$ coupling constants $f_1$ and $f_2$ could be fixed, in
principle, from the $N^*$ helicity amplitudes $A_{1/2}$ and
$A_{3/2}$~\cite{yongprc77},
\begin{eqnarray}
A_{1/2}^{p^*} &=& \frac{e\sqrt{6}}{12}\sqrt{\frac{k_{\gamma}}{M_N
M_{N^*}}} \nonumber \\
&& \left( f_1+\frac{f_2}{4M^2_N}M_{N^*}(M_{N^*}+M_N) \right ), \\
A_{3/2}^{p^*} &=& \frac{e\sqrt{2}}{4M_N}\sqrt{\frac{k_{\gamma}
M_{N^*}}{M_N}} \nonumber \\
&& \left( f_1+\frac{f_2}{4M_N}(M_{N^*}+M_N) \right ),
\end{eqnarray}
where $k_{\gamma} = (M^2_{N^*}-M^2_N)/(2M_{N^*})$, and the the
superscript $p^*$ indicates the positive-charge $D_{13}$ resonance.
The values of $f_1$ and $ f_2$ (in units of the proton charge $e$)
deduced from the helicity amplitudes quoted in the PDG ~\cite{pdg2008}
are listed in the Table~\ref{tab:nstar}. Finally, the strong couplings
$g_1$, $g_2$ and the cut-off parameters $\Lambda_s=\Lambda_t$ and
$\Lambda_R$, appearing in the form factors, are free parameters in the
present calculation, and we will fit them to the $\gamma p \to K^+
\Lambda(1520)$ differential cross section data below $E_\gamma=2.4$
GeV reported in Ref.~\cite{leps2}.

\section{Numerical results and discussion} \label{sec:results}
\begin{figure}[htdp]
\includegraphics[scale=0.4]{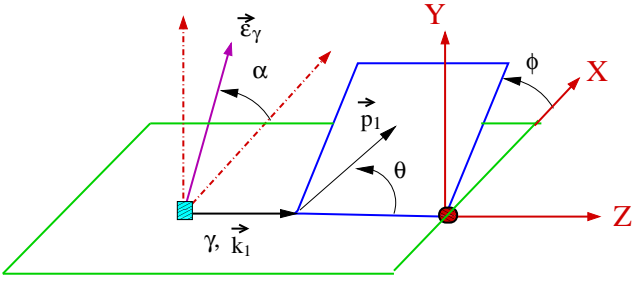}
\caption{(Color online) Definition of the different angles used
through this work. }. \label{plane}
\end{figure}

The differential cross section, in the center of mass frame (C.M.),
and for a polarized photon beam reads,
\begin{eqnarray}
\hspace{-0.2cm}\frac{d\sigma}{d\Omega}\Big|_{\rm C.M.} &=& \frac{|\vec{k}_1^{\rm
\,\, C.M.}||\vec{p}_1^{\rm \,\,C.M.}|}{4\pi^2}\frac{M_N
M_{\Lambda^*}}{(s-M_N^2)^2} \, \Big (\frac{1}{2}\sum_{s_p,
s_\Lambda^*} |T|^2 \Big ) \nonumber\\& \hspace{-0.4cm}=&
\hspace{-0.4cm} \frac{1}{2\pi}\frac{d\sigma}{d(\cos\theta_{\rm C.M.})} \left\{ 1-
\Sigma \, \cos2\left(\phi_{\rm C.M.}-\alpha\right)\right \}
 \label{eqdcs}
\end{eqnarray}
where we have taken the photon momentum in the positive $Z-$axis
direction, $\theta_{\rm C.M.}$ and $\phi_{\rm
C.M.}$ are the polar and azimuthal outgoing $K^+$ scattering angles,
$\vec{k}_1^{\rm \,\, C.M.}$ and $\vec{p}_1^{\rm \,\, C.M.}$ are the
photon and $K^+$ meson c.m. three-momenta, and the photon
polarization vector reads $\epsilon^\mu = (0, \cos \alpha, \sin
\alpha, 0)$ [see Fig.~\ref{plane}]. The asymmetry
function\footnote{Our definition of this function is consistent with
that used in previous theoretical~\cite{nam2,Nam:2006cx} and
experimental~\cite{leps1,leps2} papers. Thus, for instance and to
make contact with the definitions used in Ref.~\cite{Nam:2006cx}, we
would have there $\phi=0$, since in this reference the reaction
plane is chosen to be the $XZ$ one, while the parallel and
perpendicular directions of the photon polarization vector would
correspond to $\alpha=0$ and $\alpha=\pi/2$, respectively. }
$\Sigma$ depends on $\theta_{\rm C.M.}$, but it does not depend on
the azimuthal angle $\phi_{\rm C.M.}$.

We perform four ($g_1$,$g_2$, $\Lambda_s=\Lambda_t$ and $\Lambda_R$),
six ($g_1$,$g_2$, $\Lambda_s=\Lambda_t$, $\Lambda_R$, $M_{N^*}$ and
$\Gamma_{N^*}$ ) and eight parameter ($ef_1$, $ef_2$, $g_1$,$g_2$,
$\Lambda_s=\Lambda_t$, $\Lambda_R$, $M_{N^*}$ and $\Gamma_{N^*}$ )
$\chi^2-$fits to the LEPS $d\sigma/d(\cos\theta_{\rm C.M.})$ data at
forward angles displayed in the left panels of Fig.  2 of
Ref.~\cite{leps2}. There is a total of 59 data points. These data
correspond to forward $K^+$ angles and are given for four intervals of
$\cos\theta_{\rm C.M.}$ ranging from 1 down to 0.6. To compute the
cross sections in each interval we always use the corresponding mean
value of $\cos\theta_{\rm C.M.}$.

\begin{table*}
\caption{Values of some of the parameters used/determined in this
work. First we give the input used in each of the three fits
considered in this work: The electromagnetic couplings $ef_1$ and $ef_2$
deduced from the $N^*$ helicity amplitudes reported in the PDG (the
numerical values of the helicity amplitudes are also compiled here)
and the  mass and the total width of the $N^*$ resonance.
Next, we give results from different best fits to the $\gamma p \to K^+
\Lambda^*$ differential cross section data at forward angles and
below $E_\gamma=2.4$ GeV of Ref.~\cite{leps2}.  Finally, we also
give $\Gamma_{N^* \to \Lambda^*K}$ predicted from the parameters of
each $\chi^2-$fit, and in the case of the eight-parameter fit, we also
show the predicted $N^*$ helicity amplitudes.  }
\begin{tabular}{|c|c|c|c|}
 \multicolumn{4}{c}{Input Parameters}\\
\hline     & Fit A & Fit B & Fit C  \\
\hline  $A_{1/2}^{p^*}[\text{GeV}^{-1/2}]$ & $-0.020 \pm 0.008$ & $-0.020 \pm 0.008$  & $---$  \\
\hline  $A_{3/2}^{p^*}[\text{GeV}^{-1/2}]$ & $0.017 \pm 0.011$  & $0.017 \pm 0.011$ & $---$ \\
\hline  $ef_1$ & $0.18 \pm 0.07$  & $0.18 \pm 0.07$ & $---$ \\
\hline  $ef_2$ & $-0.19\pm 0.07$  & $-0.19\pm 0.07$ & $---$ \\
\hline  $M_{N^*}$[MeV] & $2080$  & $---$ & $---$ \\
\hline  $\Gamma_{N^*}$[MeV] & $300$  & $---$ & $---$ \\
\hline
  \multicolumn{4}{c}{Fitted Parameters}\\
\hline  $g_1$ & $5.0\pm 0.2 ^{+2.8}_{-1.5}$  & $2.0 \pm 0.1^{+1.4}_{-0.5}$ & $1.4 \pm 0.3$ \\
\hline  $g_2$ & $-9.7 \pm 2.0 ^{+6}_{-5}$  & $-3.3 \pm 0.9^{+1.8}_{-3.4}$ & $5.5 \pm 1.8$ \\
\hline  $\Lambda_s=\Lambda_t$ [MeV] & $613 \pm 2^{+5}_{-8}$   & $613 \pm 2^{+1}_{-5}$ & $604 \pm 2$ \\
\hline  $\Lambda_R$ [MeV] & $990 \pm 50 ^{+30}_{-20}$ & $5.0 \pm 3.9$~\footnote{In GeV.} & $909 \pm 55$ \\
\hline  $ef_1$ & $---$  & $---$ & $0.177 \pm 0.023$ \\
\hline  $ef_2$ & $---$  & $---$ & $-0.082 \pm 0.023$ \\
\hline  $M_{N^*}$[MeV] & $---$  & $2138 \pm 4^{+1}_{-21}$ & $2115 \pm 8$ \\
\hline  $\Gamma_{N^*}$[MeV] & $---$  & $168 \pm 10^{+19}_{-15}$ & $254 \pm 24$ \\
\hline
$\chi^2/dof$ & $2.4$  & $1.4$ & $1.2$ \\
\hline
  \multicolumn{4}{c}{Predicted Parameters} \\
\hline  $A_{1/2}^{p^*}[\text{GeV}^{-1/2}]$ & $---$ & $---$  & $0.0036 \pm 0.0086$  \\
\hline  $A_{3/2}^{p^*}[\text{GeV}^{-1/2}]$ & $---$  & $---$ & $0.058 \pm 0.021$ \\
\hline  $\Gamma_{N^* \to \Lambda^*K}$ [MeV] & $110 \pm 10 ^{+160}_{-50}$  & $43 \pm 5^{+61}_{-20}$ & $19 \pm 7$ \\
\hline  $\frac{\Gamma_{N^* \to \Lambda^*K}}{\Gamma_{N^*}}$[$\%$] & $36 \pm 3 ^{+53}_{-18}$  & $26 \pm 3^{+36}_{-12}$ & $7.5 \pm 2.8$ \\
\hline
\end{tabular} \label{tab:nstar}
\end{table*}
The fitted parameters are compiled in Table~\ref{tab:nstar}. For the
first two fits (A and B), in which the couplings $ef_1$ and $ef_2$
are fixed from the $N^*$ helicity amplitudes $A_{1/2}$ and
$A_{3/2}$, we quote two sets of errors for the fitted parameters.
The first set of errors is purely statistical and it is determined
from the the inverse of the $\chi^2-$Hessian matrix at the minimum,
while the second set accounts for some systematics of the present
approach and it takes into account the errors of $f_1$ and $f_2$
induced by the PDG helicity amplitude uncertainties.  We use a Monte
Carlo (MC) simulation to estimate these latter errors. We generate
pairs of couplings $(f_1,f_2)$ from a two-dimensional uncorrelated
Gaussian distribution with the mean values and standard deviations
quoted in Table~\ref{tab:nstar}. For each $(f_1,f_2)$ pair, we
perform a $\chi^2-$fit and thus a four dimensional distribution of
fitted parameters ($g_1$,$g_2$, $\Lambda_s=\Lambda_t$ and
$\Lambda_R$) is generated. The second set of errors, given in the
Table~\ref{tab:nstar}, accounts for the corresponding 68\%
confidence level (CL) interval deduced from that distribution.
Because its completely different origin, one can add in quadratures
the two sets of errors quoted in the Table~\ref{tab:nstar} for fits
A and B. In the case of the fit C (eight parameters), we just give
statistical errors as determined from the the inverse of the
$\chi^2-$Hessian matrix at the minimum.

\begin{center}
\begin{figure}[htdp]
\includegraphics[scale=0.4]{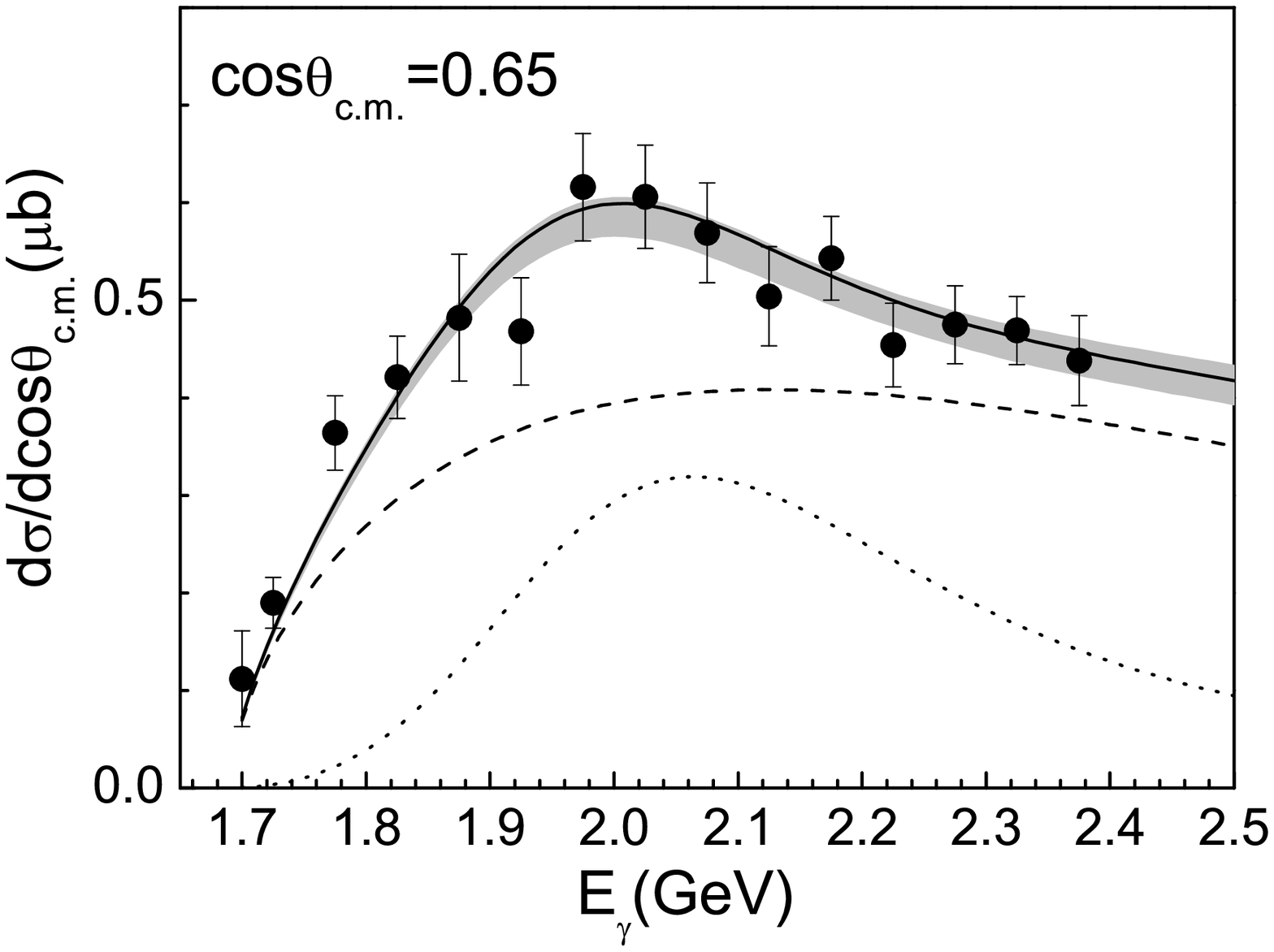} \\ %
\vspace{-0.5cm}
\includegraphics[scale=0.4]{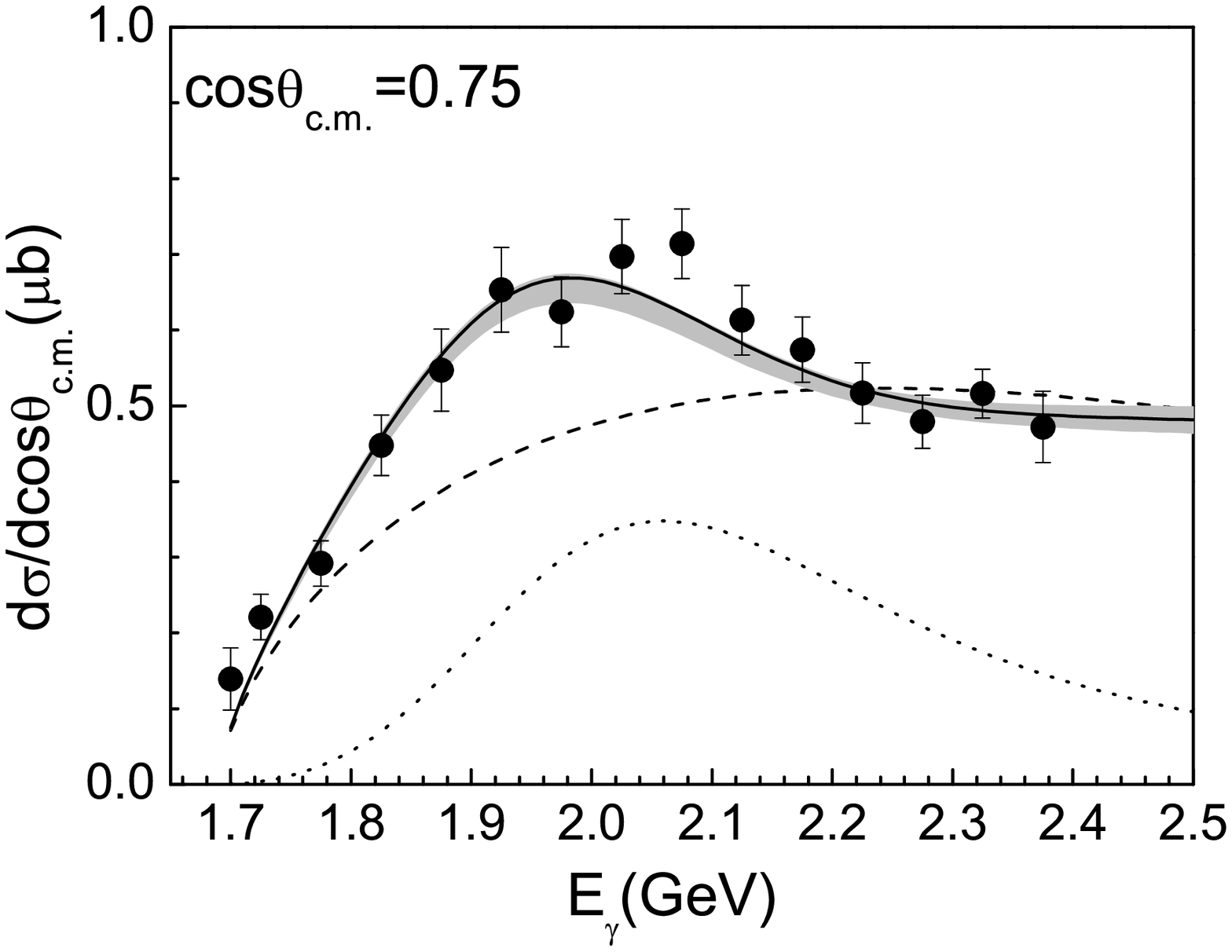} \\ %
\vspace{-0.5cm}
\includegraphics[scale=0.4]{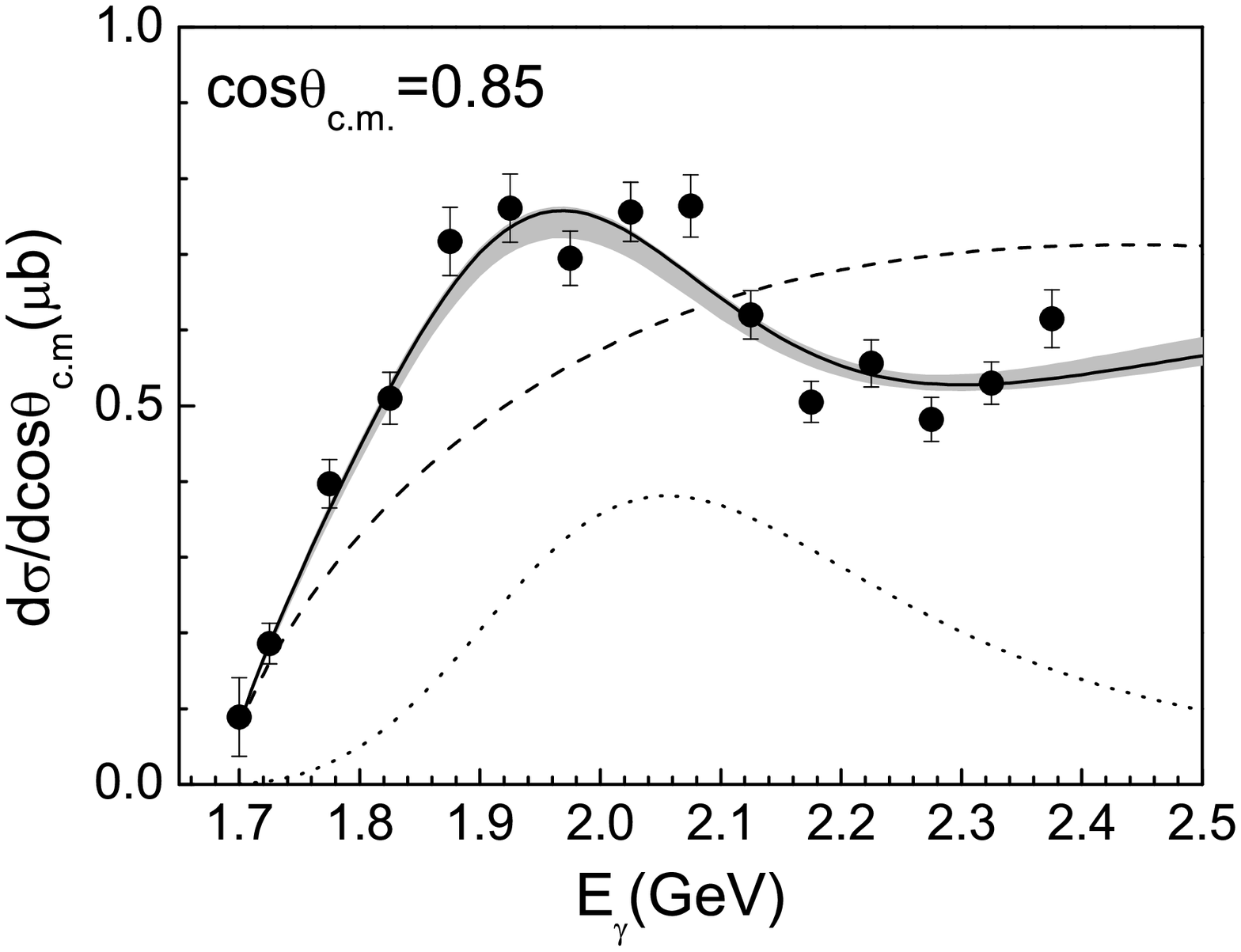} \\ %
\vspace{-0.5cm}
\includegraphics[scale=0.4]{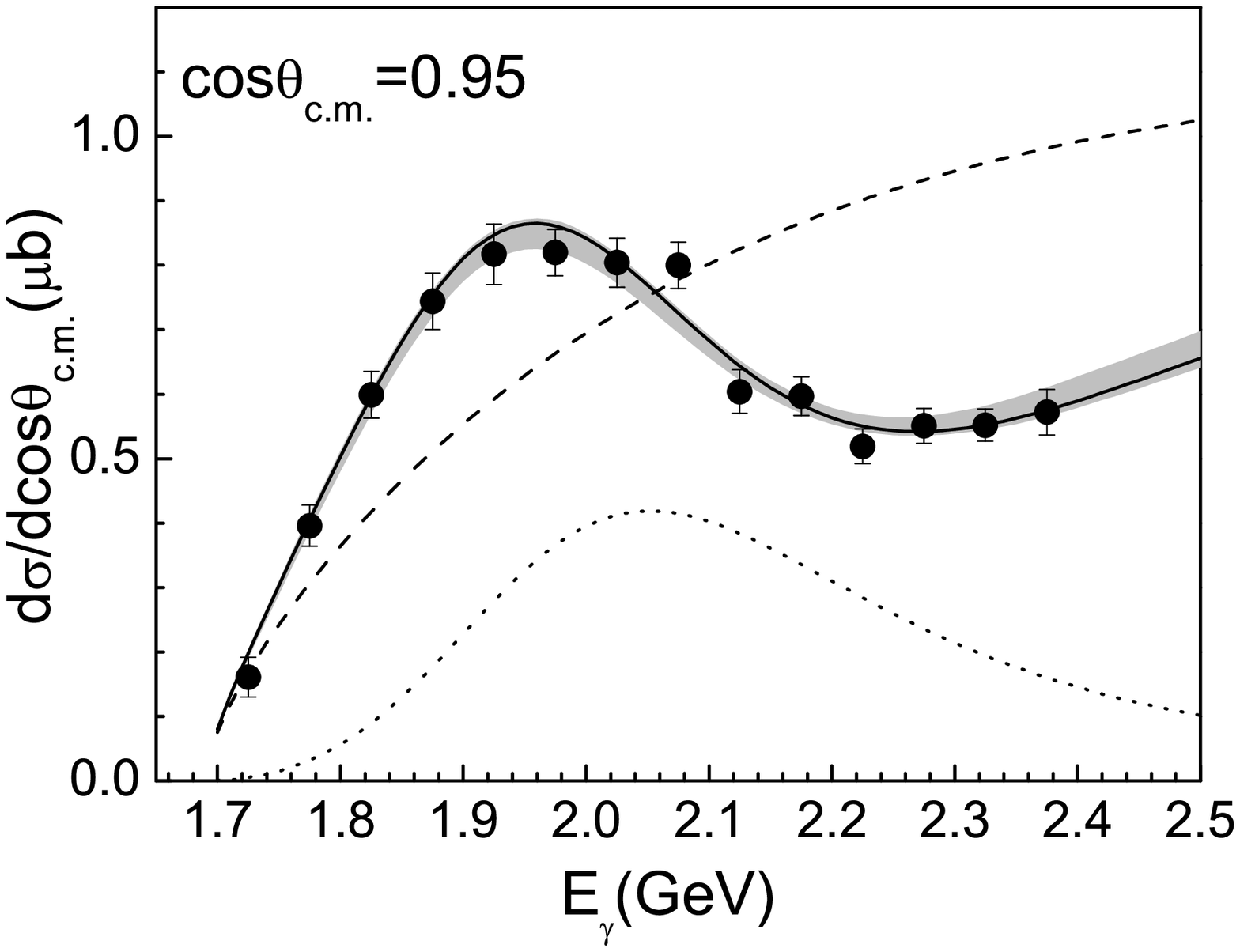}%
\caption{$\gamma p \to K^+ \Lambda^*$ differential
$d\sigma/d(\cos\theta_{\rm C.M.})$ cross sections compared with the
LEPS data~\cite{leps2}. Results have been obtained from the
eight--parameter fit C (details can be found in Table~\ref{tab:nstar}).
Dashed and dotted lines show the contributions from the background and
$N^*$ resonance terms, respectively, while the solid line displays the
full result. For this latter curve we also show the statistical 68\%
CL band.}
\label{dcs-er}%
\end{figure}
\end{center}

We find all fits displayed in the Table~\ref{tab:nstar} show
reasonable small $\chi^2/dof$. We have also performed a best fit
including only non-resonant background contributions.  It only has one
free parameter $\Lambda_s=\Lambda_t$, which turns out to be $592 \pm
1$ MeV, and  an  unacceptable $\chi^2/dof\sim 24$. Thus, the inclusion
of the nucleon resonance $N^*(2080)$ is crucial to achieve a fairly good
description of the new LEPS differential cross section data.

With the strong coupling constants obtained from the $\chi^2-$fits,
we have evaluated the $N^*(2080)$ to $\Lambda^*K$ partial decay
width,
\begin{widetext}
\begin{eqnarray}
\Gamma_{N^* \to \Lambda^*K} &=& \frac{|\vec{p}_1^{\,\,\rm C.M.}|
M_{N^*}
  (E_{\Lambda^*}-M_{\Lambda^*})}{18 \pi
  M_{\Lambda^*}^2}  \Big \{
  |\vec{p}_1^{\,\,\rm C.M.}|^4  \frac{g_2^2}{m^4_K}
+    |\vec{p}_1^{\,\,\rm C.M.}|^2(2
  E_{\Lambda^*}-M_{\Lambda^*})\frac{(M_{N^*}+M_{\Lambda^*})}{M_{N^*}}
  \frac{g_1g_2}{m_K^3} \nonumber \\
 &+& \left(\frac{M_{N^*}+M_{\Lambda^*}}{M_{N^*}}\right)^2 ( E_{\Lambda^*}^2-
  E_{\Lambda^*}M_{\Lambda^*}+ \frac52 M_{\Lambda^*}^2 ) \frac{g_1^2}{m_K^2}
  \Big \}
\end{eqnarray}
\end{widetext}
as deduced from the Lagrangian of Eq.~(\ref{eq:eqknstar}). In the
above expression $E_{\Lambda^*} = \sqrt{M_{\Lambda^*}^2 +
|\vec{p}_1^{\,\,\rm C.M.}|^2}$. The numerical predictions\footnote{We
take $M_{\Lambda^*}= 1.5195$ GeV and $m_K$= 0.4937 GeV.} for each fit
are also given in the Table~\ref{tab:nstar}. Let us first pay
attention to fit A results.  Having in mind that in that case, we have
assumed a total width of 300 MeV for the $N^*(2080)$ resonance, we
find that the $\Lambda^*K$ decay mode of this resonance will be become
the dominant one, if one attributes the observed bump structure at
forward angles, reported in the SPring-8 LEPS experiment, to the
effects produced by this resonance, as implicitly assumed in this
work. This large coupling of the two-star $D-$wave $J^P=3/2^-$
$N^*(2080)$ resonance to the $\Lambda^* K^+$ channel will confirm/get
support from the QM results of Simon Capstick, and W. Roberts in
Ref.~\cite{simonprd58}, as mentioned above. We have taken advantage of
the apparent important role played by the resonant contribution to
explain the SPring-8 LEPS $\vec{\gamma} p \to K^+ \Lambda(1520)$ data,
and we have used it to improve our knowledge on some $N^*(2080)$
properties. First, keeping the electromagnetic $ef_1$ and $ef_2$
couplings fixed, we have taken the mass and width of the resonance as
adjustable parameters. From our six-parameter fit B, we find $2138 \pm
4$ MeV and $168 \pm 10$ MeV for the resonance mass and width,
respectively (we do not quote here the errors induced by the helicity
amplitude uncertainties), and a significant improvement\footnote{Note
that for fit B, we find an unrealistic central value of 5 GeV for the
cutoff  $\Lambda_R$, with a large error ($\sim 4$ GeV), which
indicates that the $\chi^2$ is rather insensitive to this parameter.
Indeed, we get equivalent fits ($\chi^2/dof=1.4-1.6$) as long as
$\Lambda_R \ge 1$ GeV. The rest of parameters change within their
respective errors to accommodate the minor modifications induced by the
change in the resonance cut off parameter. } of the $\chi^2/dof$,
which is now of the order of 1.4. The major evidences quoted in the
PDG for the $N^*(2080)$ resonance come from analyses performed thirty
years ago~\cite{Hohler:1979yr, Cutkosky:1980rh}, and that certainly
are not inconsistent with values for the mass and width in the range
of 2140 MeV and 170 MeV, respectively. We have also explored the
possibility of determining the the electromagnetic $ef_1$ and $ef_2$
couplings of this resonance, and hence we have carried out an
eight--parameter fit (C), where these two couplings are also adjusted
to data. The $\chi^2/dof$ slightly lowers down to 1.2, being $ef_1$
and $ef_2$ nearly compatible, within errors, with those deduced form
the helicity amplitudes. Note, that these latter ones were also
reported almost 30 years ago~\cite{Awaji:1981zj}, and they might be
also subject of large uncertainties. Besides, in this fit, the $g_1$
coupling turns out to be significantly smaller than that obtained in
the first fit, what leads to a much smaller $\Gamma_{N^* \to
\Lambda^*K}/\Gamma_{N^*}$ branching fraction, probably more reasonable
than that deduced from the four parameter fit A. We conclude, that
this latter fits leads to an overall good description of data and that
it can be used to constrain some of the properties of the $N^*(2080)$
resonance.  Indeed, this can be seen in Fig.~\ref{dcs-er}, where
differential cross sections, deduced from the results of the
eight--parameter fit C, are shown and compared to data . In this
figure, dashed and dotted lines show the contributions from the
background and $N^*$ resonance terms, respectively, while the solid
line displays the full result. For this latter curve we also show the
68\% CL band obtained from the statistical uncertainties of the fitted
parameters.  We see that the bump structure in the differential cross
section at forward $K^+$ angles is fairly well described ($\chi^2/dof
\sim 1.2$) thanks to a significant contribution from the $N^*$.
\begin{center}
\begin{figure}[htdp]
\includegraphics[scale=0.4]{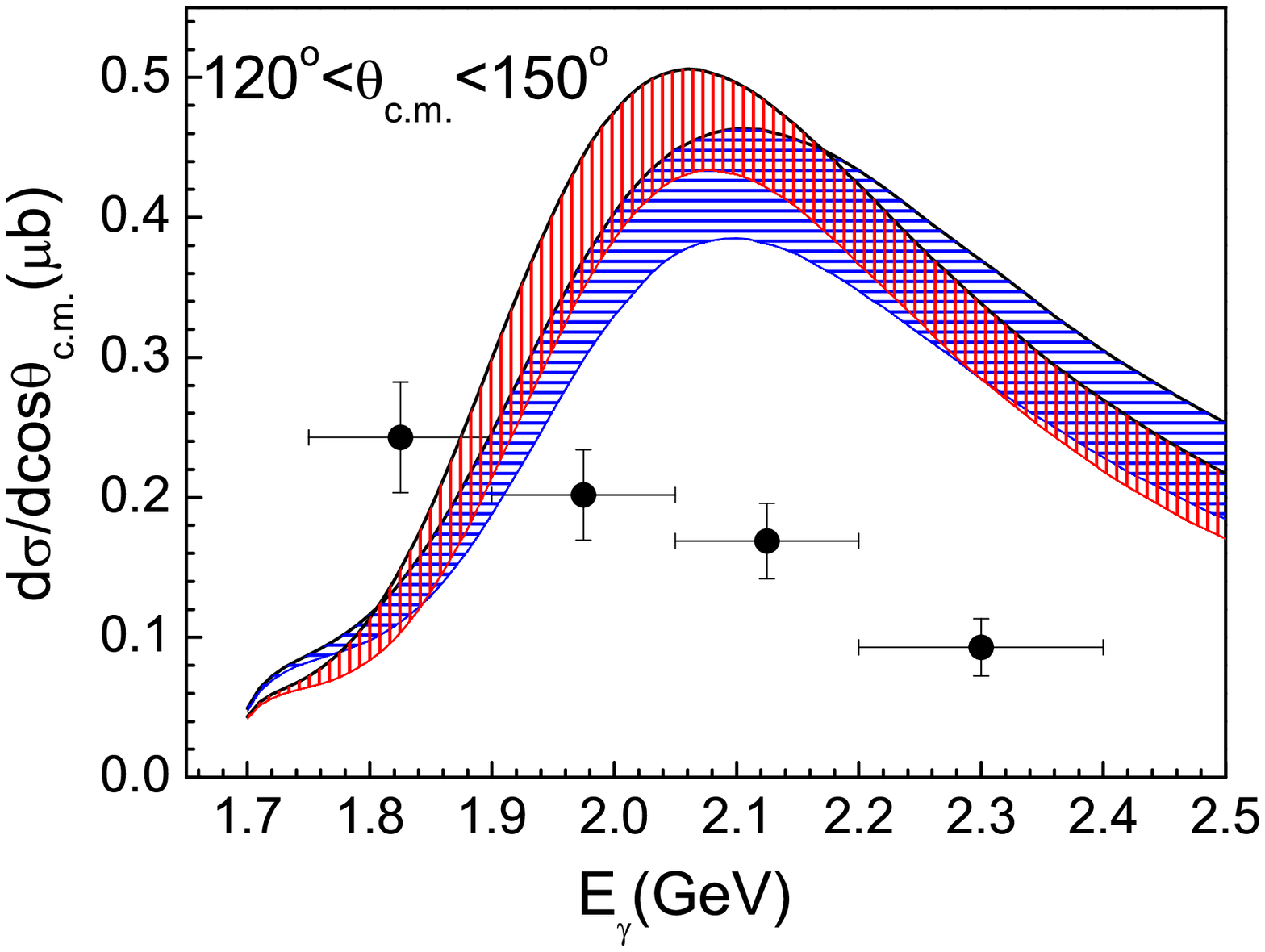} \\ %
\vspace{-0.5cm}
\includegraphics[scale=0.4]{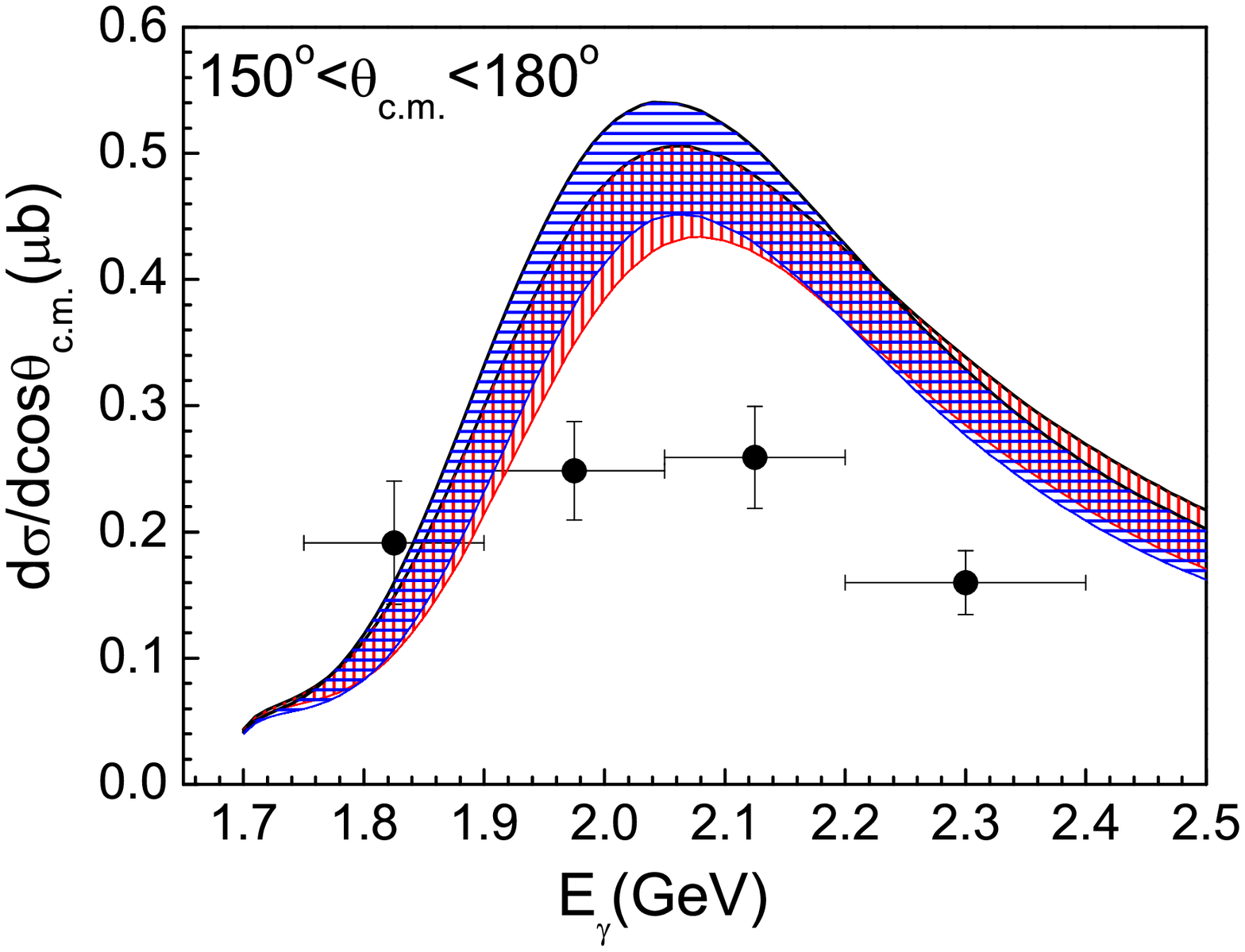} \\ %
\caption{(Color online)  $\gamma p \to K^+ \Lambda^*$ differential
$d\sigma/d(\cos\theta_{\rm C.M.})$ cross section compared with the
LEPS data~\cite{leps1}.  In both panels, we display our predicted (fit C)
68\% CL bands for the two extreme values of the corresponding
$\theta_{\rm C.M.}$ interval.  The blue horizontal
 line shaded region stand for the 68\% CL band associated
to 120$^0$,  in the upper panel and to 180$^0$
in the lower panel. In both panels, the red vertical line
shaded region stand for the 68\% CL band that corresponds to 150$^0$.}
\label{dcs-deg}%
\end{figure}
\end{center}

\begin{center}
\begin{figure}[htdp]
\includegraphics[scale=0.4]{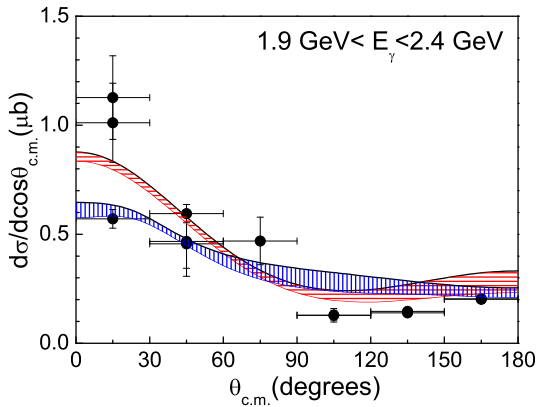}%
\caption{(Color online) $\gamma p \to K^+ \Lambda^*$ differential
$d\sigma/d(\cos\theta_{\rm C.M.})$ cross section compared with the
LEPS data~\cite{leps1}.  Shaded regions stand for our predicted (fit
C)  68\%
CL bands for 1.9 (red horizontal lines) and 2.4 (blue vertical lines) GeV
energy photons. } \label{dcsbis-deg}%
\end{figure}
\end{center}

Next, we pay attention to backward angles and in Fig.~\ref{dcs-deg}
we depict differential cross sections for large kaon scattering
angles and obtained with the fitted (eight) parameters given in the
Table~\ref{tab:nstar}, as a function of $E_\gamma$ and the kaon
scattering angle $\theta_{\rm C.M.}$. The experimental data-points
are taken from Ref.~\cite{leps1}, where events were accumulated for
two angular intervals $\theta_{\rm C.M.}$ =$(120 - 150)^0$  and
$\theta_{\rm C.M.}$ =$(150 - 180)^0$, with the photon energy
varying in the region $1.9\leq E_\gamma \leq 2.4$ GeV. In both
panels, we have computed our predictions for the two extreme values
of the corresponding $\theta_{\rm C.M.}$ interval. The shaded
regions accounts for the uncertainties inherited from those
affecting the parameters compiled in Table~\ref{tab:nstar}. The 68\%
CL error bands have been obtained using a MC simulation and they
turn out to be bigger than at forward angles because the backward
differential cross section is largely dominated by the $N^*(2080)$
pole contribution. The full polar angular dependence of the
theoretical differential cross section is compared with the data of
Ref.~\cite{leps1} in Fig.~\ref{dcsbis-deg}. To increase the
statistics,  data have been integrated over the photon
energy interval $1.9\leq E_\gamma \leq 2.4$ GeV. We have considered
photons of 1.9 and 2.4 GeV of energy and computed  the 68\% CL error
bands for each energy. Both in Fig.~\ref{dcs-deg} and
~\ref{dcsbis-deg}, we see that our theoretical model leads to
reasonable descriptions of the data.

\begin{center}
\begin{figure}[htdp]
\includegraphics[scale=0.4]{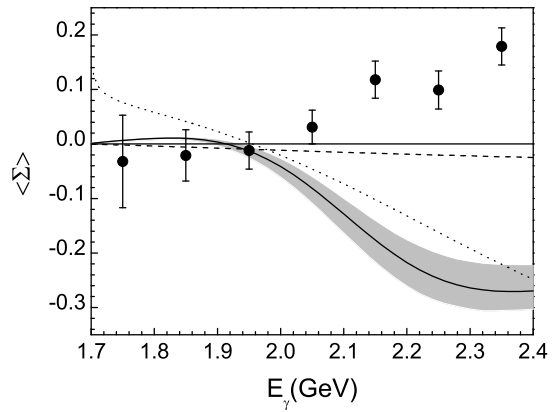}%
\caption{Polar angle ($0.6 < \cos\theta_\text {C.M.} < 1.0$) average
photon-beam asymmetry $\langle \Sigma \rangle$ as a function of
$E_\gamma$ for the $\gamma p \to K^+ \Lambda^*$ reaction. Dashed and
dotted lines show the average asymmetry from the background and $N^*$
resonance terms alone, respectively, while the solid line displays the
full result (eight--parameter fit C). For this latter curve we also show
the 68\% CL grey band obtained from the statistical errors of the
fitted parameters in Table~\ref{tab:nstar}. Data have been taken from
Ref~\cite{leps2}.}
\label{asymmetry}%
\end{figure}
\end{center}

Finally, in Fig.~\ref{asymmetry}, we compare our predictions (fit C) for the
polar angle average photon-beam asymmetry $\langle\Sigma \rangle$ as
a function of $E_{\gamma}$ with the recent SPring-8 LEPS data of
Ref.~\cite{leps2}.  We calculate $\langle \Sigma \rangle$ as
\begin{equation}
\langle \Sigma  \rangle  = \frac{\int^{1.0}_{0.6}
  \frac{d\sigma}{d(\cos\theta_{\rm C.M.})} \Sigma
(\cos \theta_\text {C.M.},E_{\gamma}) d(\cos\theta_\text
{C.M.})}{\int^{1.0}_{0.6}
  \frac{d\sigma}{d(\cos\theta_{\rm C.M.})} d(\cos\theta_\text
{C.M.})},
\end{equation}
where $\Sigma (\cos \theta_\text {C.M.},E_{\gamma})$ is defined in
Eq.~(\ref{eqdcs}). In this figure, dashed and dotted lines show the
polar angle average asymmetry from the background and $N^*$ resonance
terms alone, respectively, while the solid line displays the full
result. For this latter curve we also show the 68\% CL grey band
obtained from the statistical errors of the fitted parameters in
Table~\ref{tab:nstar}. The description of our model of this observable
is much poorer than in the rest of cases examined above, and we find
here discrepancies of about two standard deviations with data when the
theoretical uncertainties are also taken into account.  One might
think that the inclusion of a $t-$channel $K^*$ exchange might improve
the situation, since it leads to positive values for the
asymmetry~\cite{Nam:2006cx}. We have explored such a
possibility\footnote{We have used Eq.~(15) of Ref.~\cite{toki}.}, but
we have found tiny changes even for values of the $K^*\Lambda^*N$
coupling constant as large as 10, strongly disfavored by the
theoretical findings of Ref.~\cite{toki}, and certainly the inclusion
of this mechanism would not significantly improve the situation. We
have also studied the effect of including a relative complex phase
between the background and the $N^*$ contributions. With values of
this phase around 130--140$^0$ and re-adjusting the rest of the
parameters, we find values of $\langle \Sigma \rangle $, though still
negative, much smaller (in absolute values) and closer to zero than
those displayed in Fig.~\ref{asymmetry}. Moreover, the fair agreement
with the $d\sigma/d\cos\theta_{\rm C.M.}$ data exhibited in
Fig.~\ref{dcs-er} is not spoiled out, at all.

We would like to point out that if the sign of our calculated
asymmetry is changed, we would find a better description of data.
However, in the best of our knowledge our definition of the
asymmetry is consistent with that stated in Ref.~\cite{leps2}, and
given in terms of the ratio $(N_v-N_h)/(N_v+N_h)$, where $N_v$ and
$N_h$ are the $\Lambda^*$ yields with vertically and horizontally
polarized photons, respectively. Yet, we would like to mention that
the authors of Ref.~\cite{nam3} also find negative values for photon
beam asymmetry, as can be seen in Fig. 9a of that reference.

\section{Summary and Conclusions } \label{sec:conclusions}

We have studied the $\vec{\gamma} p \to \Lambda^*K^+$ reaction at low
energies within a effective Lagrangian approach. In particular, we
have paid an special attention to a bump structure in the differential
cross section at forward $K^+$ angles reported in the recent SPring-8
LEPS experiment~\cite{leps2}. Starting from the background
contributions studied in previous works, we have shown that this bump
might be described thanks to the inclusion of the nucleon resonance
$N^*(2080)$ (spin-parity $J^P = 3/2^-$). We have fitted its mass,
width and hadronic $\Lambda^*K^+$ and electromagnetic $ N^* N\gamma $
couplings to data. We have found that this resonance would have a
large decay width into $\Lambda^*K$, which will be compatible with the
findings of the QM approach of Ref.~\cite{simonprd58}. We have also
calculated differential cross sections at backward angles and the
polar angle average photon-beam asymmetry. In the first case, our
results compare reasonably well with data, while for the case of the
photon-beam asymmetry the agreement with the experimental measurements
is much poorer, and we find discrepancies of about two standard
deviations for photon energies above 2 GeV. The proposed scheme here,
should be supplemented with some other reaction mechanisms which could
improve the achieved description of the photon-beam asymmetry data.

Other explanations of the observed bump in the SPring-8 LEPS data
are also possible. Indeed, in the very same experimental paper
(Ref.~\cite{leps2}) where the data is published, it is suggested
that this structure might be due to a $J^P=\frac{3}{2}^+$ nucleon
resonance, with a mass of $2.11$ GeV and a width of $140$ MeV.
However, a nucleon resonance with these features is not listed in
the PDG book~\cite{pdg2008}. In Ref.~\cite{leps2}, it is also
mentioned the possibility of a sizeable contribution from a higher
($J^P=\frac{5}{2}^-$) baryon state and/or the existence of a new
reaction process, for instance, an interference with $\phi$
photo-production~\cite{Hosaka,alvin}.

However, we have shown here that the photo-production of the $N^*(2080)$
resonance off the proton and its subsequent decay into $\Lambda^*K^+$
might also provide a simple explanation of the bump structure observed
in the experimental data.  This contradicts the findings of
Refs.~\cite{nam2,nam3}, where unnecessarily small $N^*(2080)\Lambda^*
K^+$ couplings and probably a too large width for this resonance were
used.

Finally, we would like to stress that thanks to the important role
played by the resonant contribution in the $\vec{\gamma} p \to K^+
\Lambda(1520)$ reaction, accurate data for this reaction can be used
to improve our knowledge on some $N^*(2080)$ properties, which are at
present poorly known. This work
constitutes a first step in this direction.

\section*{Acknowledgments}
We warmly thank M.J. Vicente--Vacas and J. Martin Camalich for
useful discussions. This work is partly supported by DGI and FEDER
funds, under contract FIS2008-01143/FIS, the Spanish
Ingenio-Consolider 2010 Program CPAN (CSD2007-00042), and
Generalitat Valenciana under contract PROMETEO/2009/0090. We
acknowledge the support of the European Community-Research
Infrastructure Integrating Activity "Study of Strongly Interacting
Matter" (acronym HadronPhysics2, Grant Agreement n. 227431) under
the Seventh Framework Programme of EU. Work supported in part by DFG
(SFB/TR 16, "Subnuclear Structure of Matter"). Ju-Jun Xie
acknowledges Ministerio de Educaci\'on Grant SAB2009-0116.

\end{document}